\documentclass{article}
\usepackage{helvet}
\usepackage{courier}
\usepackage[latin9]{inputenc}
\usepackage[letterpaper]{geometry}
\usepackage{color}
\usepackage{prettyref}
\usepackage{float}
\usepackage{amsmath}
\usepackage{graphicx}
\usepackage[numbers]{natbib}

\makeatletter

\providecommand{\tabularnewline}{\\}
\newcommand{\lyxdot}{.}

\floatstyle{ruled}
\newfloat{algorithm}{tbp}{loa}
\providecommand{\algorithmname}{Algorithm}
\floatname{algorithm}{\protect\algorithmname}

\def\year{2017}\relax
\usepackage{times}

\floatstyle{ruled}
\newfloat{algorithm}{tbp}{loa}
\providecommand{\algorithmname}{Algorithm}
\floatname{algorithm}{\protect\algorithmname}

\makeatother

\begin{document}
\global\long\def\sps{\mathrm{\:spikes/s}}
 \global\long\def\mV{\,\mathrm{mV}}
 \global\long\def\ms{\,\mathrm{ms}}
\global\long\def\s{\,\mathrm{s}}
\global\long\def\ns{\,\mathrm{nS}}
\global\long\def\pa{\,\mathrm{pA}}

\title{Neurogenesis and multiple plasticity mechanisms enhance associative
memory retrieval in a spiking network model of the hippocampus}

\author{Yansong, Chua and Cheston, Tan\\
 Institute for Infocomm Research, A{*}STAR, Singapore\\
 chuays@i2r.a-star.edu.sg\\
 }
\maketitle
\begin{abstract}
Hippocampal CA3 is crucial for the formation of long-term associative
memory. It has a heavily recurrent connectivity, and memories are
thought to be stored as memory engrams in the CA3. However, despite
its importance for memory storage and retrieval, spiking network models
of the CA3 to date are relatively small-scale, and exist as only proof-of-concept
models. Specifically, how neurogenesis in the dentate gyrus affects
memory encoding and retrieval in the CA3 is not studied in such spiking
models. Our work is the first to develop a biologically plausible
spiking neural network model of hippocampal memory encoding and retrieval,
with at least an order-of-magnitude more neurons than previous models.
It is also the first to investigate the effect of neurogenesis on
CA3 memory encoding and retrieval. Using such a model, we first show
that a recently developed plasticity model is crucial for good encoding
and retrieval. Next, we show how neural properties related to neurogenesis
and neuronal death enhance storage and retrieval of associative memories
in the CA3. In particular, we show that without neurogenesis, increasing
number of CA3 neurons are recruited by each new memory stimulus, resulting
in a corresponding increase in inhibition and poor memory retrieval
as more memories are encoded. Neurogenesis, on the other hand, maintains
the number of CA3 neurons recruited per stimulus, and enables the
retrieval of recent memories, while forgetting the older ones. Our
model suggests that structural plasticity (provided by neurogenesis
and apoptosis) is required in the hippocampus for memory encoding
and retrieval when the network is overloaded; synaptic plasticity
alone does not suffice. The above results are obtained from an exhaustive
study in the different plasticity models and network parameters.
\end{abstract}

\section{Introduction}

\label{sect_intro}

It is well known that the brain region called the hippocampus is heavily
involved in long-term memory storage and retrieval~\citep{Squire2009}.
In particular, a sub-region of the hippocampus known as the CA3 is
crucial for the formation of long-term declarative memory, which includes
both spatial~\citep{Moser2008} and episodic~\citep{Tulving2002}
memory. CA3 has heavily recurrent connectivity (i.e.\ CA3 neurons
form many synaptic connections with other CA3 neurons), more so than
any other brain area~\citep{Rolls1997}, and memories are thought
to be stored as pattern of synaptic weights in the CA3, typically
referred to as memory engrams.

However, despite the well-known importance of the CA3 for memory storage
and retrieval, to date, spiking neural network models studying such
a role exist as only small-scale models. These models, while useful
for illustrating certain computational principles, contain only up
to 30 CA3 neurons (e.g.~\citep{Brijesh2007,Tan2013}). Moreover,
they lack biological realism. There is hence a need for developing
a computational model of the CA3 utilizing spiking neural networks
to facilitate further studies of its functionality. Such a model should
be organized as a recurrent network, with mutually connected excitatory
and inhibitory populations, whilst the neurons are effectively in
the fluctuation-driven regime, with near threshold mean membrane potential.

Our work is the first to develop a biologically realistic spiking
neural network model of hippocampal memory encoding and retrieval,
with over two orders-of-magnitude as many neurons in the CA3 compared
to previous models. It is also only a recent development whereby a
combination of different plasticity models (using spike timing dependent
plasticity, STDP) acting on different time-scales are shown to form
long-lasting memory engrams in a biologically realistic spiking neural
network; while previous works have consistently failed to form long-lasting
memory engrams of a reasonable number without interference~\citep{Chrol-Cannon15,Kunkel11_00160}.

Other studies have used feedforward, rate-based models of CA3, which
are less realistic. It is therefore not clear how well their findings
generalize to recurrent spiking networks, which the CA3 is one such
network. Such networks have very different properties from simpler
networks, and one major challenge is their stability~\citep{Zenke2015,Abbott2000}.
Neurogenesis (neuron birth) and neural apoptosis (neuron death) in
the dentate gyrus add further complexity, whose role in memory formation
in the CA3 has yet to be investigated in realistic models. We further
note that among models of neurogenesis, most use rate-based neurons,
with~\citep{Aimone2009} the only exception (using spiking neurons).
However, they do not model the CA3 region.

In~\citep{Zenke2015}, new plasticity mechanisms are developed to
enable stable encoding of memories in plastic recurrent networks that
can be retrieved hours (network time) later. However, the plasticity
mechanisms have yet to be shown to be reliable for memories of a realistic
number. Also, it is not clear what role neurogenesis plays in memory
formation, given a plastic spiking network encoding a reasonable number
of memories. Thus, our paper is the only work that investigates the
effect of neurogenesis on a plastic spiking model of CA3.

Using our model, we first show that a novel plasticity rule~\citep{Zenke2015}
is crucial for good encoding and retrieval. Next, we show how properties
related to neurogenesis (i.e. increased excitability of newborn neurons)
in the dentate gyrus enhances memory encoding and retrieval in the
CA3, while keeping the dentate gyrus population to a constant size
(neuronal death is simulated whereby neurons are recycled to encode
new memories). From our model, we discover that with increasing number
of memories encoded, the network without neurogenesis fails to retrieve
any of the previously encoded memories, while neurogenesis allows
the more recent memories to be retrieved. 

\section{Methods}

\label{sect_methods}

To study encoding and retrieval of associative memories in the hippocampus,
we construct a three-layer spiking neuron network model of the hippocampus
(Fig.~\ref{fig:Network-setup}). The three layers are respectively
the entorhinal cortex (EC), the dentate gyrus (DG) and CA3. For simplicity,
we do not model CA1, which is thought to perform a simple comparator
or decoding role~\citep{Tulving2002,Rolls1997,rolls1987information}.
Incoming memories are simulated by stimulating a randomly selected
subset of the EC neurons, which would trigger spiking activity that
cascades to the other two layers via different pathways. Using three
different sets of STDP models (see \ref{sect_plasticity_models}),
we compare which of these would enable the stable encoding and retrieval
of these memories. In this section, we describe first the hippocampal
network setup, next the various plasticity rules, the different simulation
settings and finally the measures to quantify memory retrieval for
our study.

\subsection{Network setup}

\begin{figure}[h]
\begin{centering}
\includegraphics[width=0.4\textwidth]{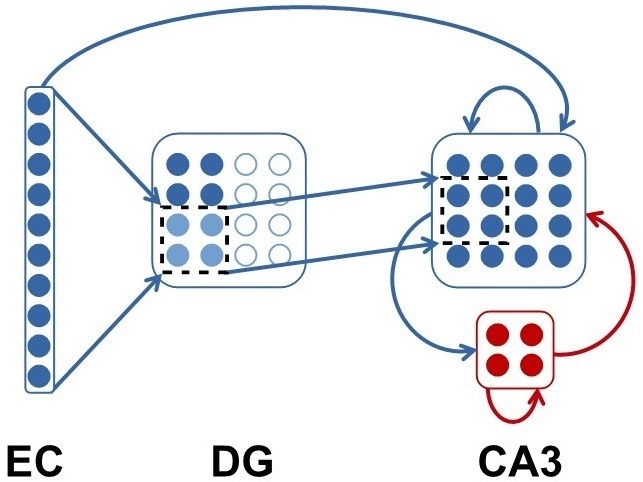}
\par\end{centering}
\caption{Network setup. Blue circles denote excitatory neurons and red circles
denote inhibitory neurons. Light blue circles and unfilled circles
denote newborn and yet-to-be-born DG neurons respectively. Blue arrows
denote excitatory synapses and red arrows denote inhibitory synapses.\label{fig:Network-setup}}
\end{figure}

The hippocampal network is made up of adaptive integrate-and-fire
(AIF) neurons~\citep{Zenke2015}. The entorhinal cortex is modeled
using $1024$ standalone AIF neurons, which receive no other inputs
other than the memory stimuli. The DG is modeled using standalone
AIF neurons. Depending on the simulation settings, the size of the
DG network can be $3200$ neurons. The CA3 network is a recurrent
one, comprising of $3200$ excitatory neurons and $800$ inhibitory
neurons. In the recurrent network, the neurons are connected to each
other with a Gaussian connectivity profile, such that neurons that
are close to each other have a higher connectivity probability. EC
neurons are connected to the CA3 excitatory neurons with a connection
probability of $0.15$. All DG and CA3 neurons receive fluctuating
Poisson inputs to achieve a low firing rate of $0.3\sps$ and a mean
membrane potential of $5\mV$ below spiking threshold. EC-CA3 connections
(with plastic weights) are formed at the beginning of each simulation,
while EC-DG and DG-CA3 connections (with static weights) are only
formed prior to each new memory to be encoded during the simulation,
to emulate structural plasticity \prettyref{fig:Network-setup}.

\subsection{Plasticity models}

\label{sect_plasticity_models}

We have three classes of plasticity models in our study, namely short
term plasticity~\citep{Tsodyks00}, STDP plasticity for excitatory
synapses and STDP plasticity for inhibitory synapses impinging on
excitatory neurons~\citep{Vogels11_1569}. All three are needed for
successful encoding and retrieval~\citep{Zenke2015}. There are three
different types of excitatory plasticity in our study, namely pair
STDP~\citep{Song00}, triplet STDP~\citep{Pfister06_9673}, and
unified plasticity, which is a simplified version of the plasticity
model described in~\citep{Zenke2015}, which incorporates Hebbian,
heterosynaptic and transmitter-induced plasticity mechanisms and was
shown to reliably form memory engrams that can be retrieved with little
interference.

In all network settings, the synaptic connections between EC-CA3 excitatory
population and CA3-CA3 excitatory populations are plastic ( with different
learning rates, using one of the three excitatory plasticity rules,
plus short-term plasticity), the CA3 excitatory to CA3 inhibitory
population synaptic connections are plastic (short-term plasticity)
and the CA3 inhibitory to CA3 excitatory population synaptic connections
are plastic (inhibitory plasticity).

The dynamics of short term plasticity together with the neuron dynamics
are described in \citep{Zenke2015}; the other plasticity mechanisms
are now described. The dynamics of pair STDP is described by: {\small{}
\[
\dot{W^{ij}}=A_{+}K^{i}S^{j}(t)-A_{-}K^{j}S^{i}(t),
\]
}whereby the time evolution of the pre-synaptic variable $K^{i}$
and postsynaptic variable $K^{j}$ are respectively described by $\dot{K^{i}}=-\frac{K^{i}}{\tau_{K^{i}}}+S^{i}(t)$
and $\dot{K^{j}}=-\frac{K^{j}}{\tau_{K^{j}}}+S^{j}(t)$ and $A_{+/-}$
are the learning rates. The dynamics of the triplet plasticity rule
is described by: {\small{}
\[
\dot{W^{ij}}=K_{1}^{i}A_{+}(1+K_{2}^{j}(t-\epsilon)S^{j}(t))-K_{1}^{j}A_{-}(1+K_{2}^{i}(t-\epsilon)S^{i}(t))
\]
}whereby synaptic variables $K_{1/2}^{i/j}$ evolve as per synaptic
variables described for paired STDP, $A_{+/-}$ are the learning rates
and $t-\epsilon$ ensures that weight update is done prior to spike
times. The unified plasticity model is described by: {\small{}
\[
{\dot{W^{ij}}=K^{i}A_{+}K_{2}^{j}(t-\epsilon)S^{j}(t)-K_{1}^{j}A_{-}S^{i}(t)\atop -\beta(W^{ij}-\tilde{W^{ij})}(K_{1}^{j}(t-\epsilon))^{3}S^{j}(t)+\delta S^{i}(t),}
\]
}whereby synaptic weight tends towards $\tilde{W^{ij}}$, and every
pre-synaptic spike induces a small increase $\delta$ in synaptic
weight. All synaptic variables $K$ and weight change amplitudes $A$
are as described earlier. The inhibitory plasticity is described by:
{\small{}
\[
\dot{W^{ij}}=A(K^{i}S^{j}(t)+(K^{j}-\alpha)S^{i}(t)),
\]
}whereby $\alpha$ is the depressing factor. The values for the different
plasticity parameters are listed in \prettyref{sec:Supplementary-materials}.

\subsection{Simulation settings}

Our simulations have two phases: the encoding phase and the retrieval
phase. During the encoding phase, $80$ out of the $3200$ DG neurons
are selected to receive output spikes from $32$, $96$ or $160$
randomly selected EC neurons which are stimulated as a new memory
to be encoded. In the non-neurogenesis setting, to simulate the case
whereby the stimulus can be encoded by any mature DG neuron, these
$80$ DG neurons are randomly selected. In the neurogenesis (NG) setting,
in the beginning, to simulate the case whereby newborn DG neurons
are solely responsible for encoding a new memory stimulus, the entire
excitatory DG population is divided into blocks of $80$, where each
block is targeted by a new memory stimulus. In addition, under the
neurogenesis setting, the newborn DG neurons are also more excitable~\citep{Finnegan2015,Aimone2009},
which are simulated by decreasing the spiking threshold of the $80$
neurons targeted by the randomly selected EC neurons for the duration
of the encoding phase of each stimulus ($500\ms$). More precisely,
by ``newborn neurons'', we are emulating the remaining newborn DG
cells after NG and subsequent cell death of a fraction of these cells
due to inactivity.

When all the available blocks of $80$ DG neurons have been assigned
(e.g. after $40$ stimuli, requiring $40\times80=3200$ DG neurons),
we ``recycle'' the DG neurons, under the assumption of ``turnover
homeostasis'', whereby the rate of neuron birth and death is relatively
balanced~\citep{Meltzer2005}, hence maintaining the network size
of the DG population. For each new stimulus to be encoded henceforth,
$80$ out of the DG neurons are randomly selected. The synaptic weights
of incoming (from EC) and outgoing (to CA3) connections are then set
to $0\ns$ (simulating neural death) in the neurogenesis case, but
retained in the non-neurogenesis case. In the NG case, the spiking
threshold is again lowered to emulate greater excitability. These
DG neurons are then used to encode the new stimulus from a group of
randomly selected EC neurons. We have simulated up to a total of $200$
memories.

To represent a memory in the EC, $32$, $96$ or $160$ out of the
$1024$ EC neurons (these numbers are chosen for completeness in simulation
settings) are randomly selected at the start of each encoding phase
to receive a memory stimulus. They are then connected to the selected
DG neurons. The selected DG neurons are then randomly connected to
a cluster of $100$ CA3 excitatory neurons ,which emulate the memory
engram ``selected'' after memory encoding in the CA3 has stabilized,
akin to the $80$ DG neurons selected per memory. The randomly selected
EC neurons are then stimulated with $50$ spikes each over a duration
of $250\ms$. Spiking activities would then cascade down to the CA3
region via the EC-CA3 pathway and the newly formed EC-DG-CA3 pathway.
The encoding phase involves stimulating the network with up to $200$
memory stimuli, each lasting for $500\ms$ ($250\ms$ of stimulation,
then letting the network activity settle for another $250\ms$).

After all stimuli have been encoded, they are then retrieved in chronological
order. The retrieval process is very similar to encoding (e.g. same
set of $32$, $96$ or $160$ EC neurons for each memory is stimulated
in the same way), except that during retrieval, only the EC-CA3 pathway
is active, while the EC-DG-CA3 pathway is inactive (synaptic weight
set to $0\ns$), a phenomenon observed biologically and common in
models~\citep{Becker2009,Weisz2009,Hasselmo04_207}.

In all, there are two aspects in which different simulation settings
are investigated and vary, namely 1) the different excitatory plasticity
models (pair STDP, triplet STDP, unified plasticity), and 2) how DG
neurons are selected to encode memories (with or without neurogenesis).

The network is simulated with a range of different learning rates
on the two excitatory plastic connections: EC-CA3 (feed-forward, FF)
and CA3-CA3 (recurrent, REC), so as to investigate the robustness
of the results. Network parameters are given in \prettyref{sec:Supplementary-materials}.
All simulations are done on the NEST simulator~\citep{Gewaltig12}.

\subsection{Measures: Signal-to-Noise Ratio (SNR)}

After a simulation with for instance $40$ memories to be encoded
and retrieved, we next proceed to analyze the quality of the retrieved
memory engrams in the CA3. To quantify how well each encoded memory
is retrieved, a signal-to-noise ratio ($SNR$) measure is devised.
For each simulation, we know precisely when the stimulus for each
memory to be encoded is introduced into the network. Likewise for
the stimulus to retrieve an encoded memory. We also know which set
of CA3 excitatory neurons are stimulated during encoding in the time
window of $500\ms$, which we denote as $N_{enc}$. For each memory
retrieval, we systematically vary the size of the set of retrieved
CA3 excitatory neurons $N_{ret}$ by scanning through a range of spike
counts, and compute its corresponding $SNR$ (see \prettyref{alg:SNR})
for each retrieval. After this has been done for all memory retrievals,
we pick the maximum $SNR$, based on a single value for spike count
for CA3 excitatory neurons across all $40$ retrievals, and note down
the mean size of $N_{ret}$ across all $40$ retrievals.

\begin{algorithm}[h]
{\small{}$\forall A\in N_{ret}$,}{\small \par}

{\small{}$\ \ \ \ $If $A\in N_{enc}$, $Count_{signal}+1$}{\small \par}

{\small{}$\ \ \ \ $else, $Count_{noise}+1$}{\small \par}

{\small{}$SNR=\frac{Count_{signal}-Count_{noise}+|N_{enc}|}{2*|N_{enc}|}$}{\small \par}

\caption{Computation of SNR\label{alg:SNR}}
\end{algorithm}

The $SNR$ is normalized to an interval of $[0,1]$, since it is only
in the extreme cases (such as network spiking activities are highly
synchronized and persistent due to high learning rates in the REC
plasticity) that $Count_{noise}>|N_{enc}|$, given that the network
is in a fluctuation driven regime with low firing rate of $\approx0.3\sps$.
Hence the $SNR$ is unlikely to go below $0$, as confirmed by our
data collected. Therefore, a score of $0$ most likely means that
$N_{ret}\cap N_{enc}=\emptyset$, a score of $1$ must mean that $N_{ret}=N_{enc}$,
and a score of $0.5$ must mean that $Count_{signal}=Count_{noise}$,
whereby there are as many signal neurons as there are noise neurons
in $N_{ret}$ (i.e. retrieval is as good as random), one special case
being $N_{ret}=\emptyset$.

\section{Results}

\label{sect_results}

In this section, we first present results for the comparison of all
$3$ plasticity rules using single memory encoding and retrieval within
a certain parameter range of feedforward and recurrent plasticity
learning rates, and show that the unified plasticity model has the
best retrieval results consistently. After which, using only the full
plasticity model, we further justify that the range of learning rates
can be narrowed down, and switched from a linear to logarithmic scale
based on simulation results of memory encoding and retrieval of $3$
stimuli. Third, using $40$ stimuli (hence fully utilizing the entire
dentate gyrus population), we explore the full parameter space (of
different learning rates, number of EC neurons per stimulus, with
and without NG) and determine the parameters that give the best $SNR$.
In particular we investigate under what parameter space does neurogenesis
enables better memory retrieval compared to without neurogenesis.
Fourth, as the $SNR$ results in the third set of simulations are
not perfect (i.e. $1$), we investigate the sources of noise in the
network, and study how each of them affects memory retrieval. In all,
$4$ noise sources are identified and their effect on memory retrieval
are systematically investigated. Finally, using learning rates with
the best $SNR$ (for both FF plasticity only, and FF+REC plasticity;
and EC neurons per stimulus$=\{32,96\}$) from the third set of simulations,
we investigate how the network would perform for memory retrieval
of up to $200$ memories ($5$ times overloading of the dentate gyrus).
Retrieval of $200$ memories is also conducted using only partial
retrieval such that only a fraction of the encoding EC neurons are
stimulated during retrieval. For comparison, the same set of simulations
are re-run but without NG. The purpose of the above simulations is
to investigate whether NG improves $SNR$ results, and how it interacts
with FF and REC plasticity. For large number of memories ($200$),
we show that NG enables the retrieval of recent memories while without
NG, memory retrieval is as good as random. We further investigate
the underlying mechanisms behind this observation. The results presented
below are for the case with NG unless otherwise stated.

\subsection*{Comparison of plasticity models\label{subsec:Comparison-of-plasticity}}

\begin{figure}
\begin{centering}
\includegraphics[width=0.5\textwidth]{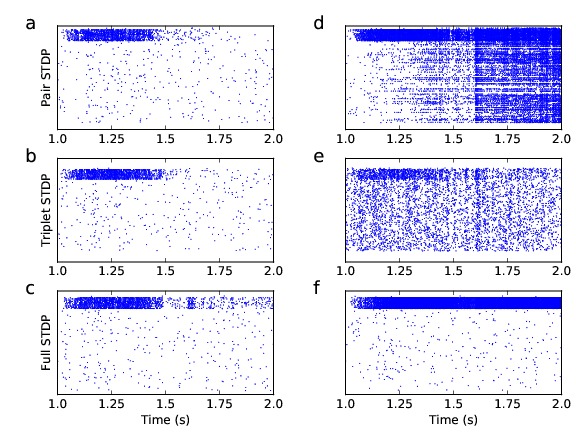}
\par\end{centering}
\caption{Raster plots for encoding (at $1\protect\s$) and retrieval (at $1.5\protect\s$)
of a single stimulus with different plasticity models. (a-c) Raster
plots for simulations without NG, $32$ EC neurons, and with FF and
REC learning rates at $0.01$. (d-f) Raster plots for simulations
with NG, $160$ EC neurons, and with FF and REC learning rates at
$0.15$. \label{fig:3pl}}
\end{figure}

\begin{figure}
\begin{centering}
\includegraphics[scale=0.5]{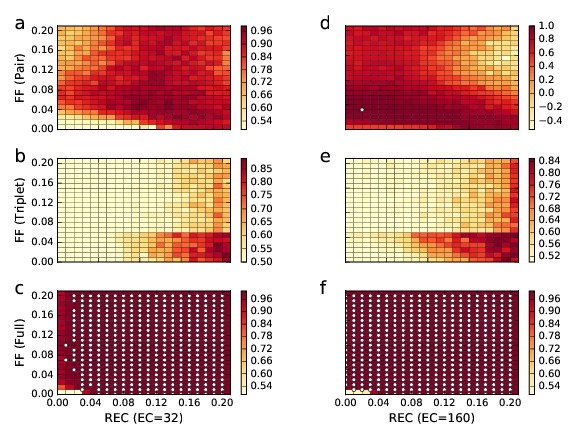}
\par\end{centering}
\caption{Colormaps for encoding (at $1\protect\s$) and retrieval (at $1.5\protect\s$)
of a single stimulus with different plasticity models. (a-c) Colormaps
for $32$ EC neurons per stimulus, with pair, triplet and full plasticity
models respectively. White markers denote learning rates with $SNR=1$
(d-f) As per (a-c), with $160$ EC neurons per stimulus.\label{fig:Colormaps-for-encoding}}

\end{figure}

In this first set of simulations, we investigate across a set of parameters
(number of EC neurons per stimulus, with and without NG, range of
FF and REC learning rates), how the $3$ different plasticity models
compare using $SNR$ as a metric. We use a range of learning rates,
from $0.0$ to $0.2$, with an interval of $0.01$. From \prettyref{fig:3pl},
we observe that the full plasticity model, at low learning rates and
$32$ EC neurons per stimulus, is able to trigger sufficient spiking
activities during retrieval (\prettyref{fig:3pl}c), as opposed to
the other $2$ plasticity models (\prettyref{fig:3pl}a,b). At higher
learning rates and $160$ EC neurons per stimulus, the pair and triplet
plasticity models are also able to retrieve the memory encoded to
some extent but in the process generate considerable amount of background
noise such that their $SNR$ suffer (see \prettyref{fig:3pl}d,e).
On the other hand, the full plasticity model is able to retrieve the
encoded memory without generating much background noise (see \prettyref{fig:3pl}f).
This demonstrates how the homeostatic terms in the model operate on
a similar time-scale as the STDP terms and help to balance network
activities. However, spiking activities from memory encoding last
beyond the usual time window of $500\ms$ and continue into the retrieval
time window. Hence, from this set of simulations with $3$ different
plasticity models, we confirm that the full plasticity models outperform
the other $2$ models in memory encoding and retrieval across a broad
range of parameters (\prettyref{fig:Colormaps-for-encoding}), but
persistent activities at high learning rates (present in all $3$
plasticity models) can become a problem for simulation with multiple
memories. We investigate this in the next set of simulations using
$3$ stimuli, whilst using the same parameters. All subsequent simulations
are performed using the full plasticity model.

\subsection*{Defining parameter space for learning rates\label{subsec:Defining-parameter-space} }

\begin{figure}
\begin{centering}
\includegraphics[width=0.5\textwidth]{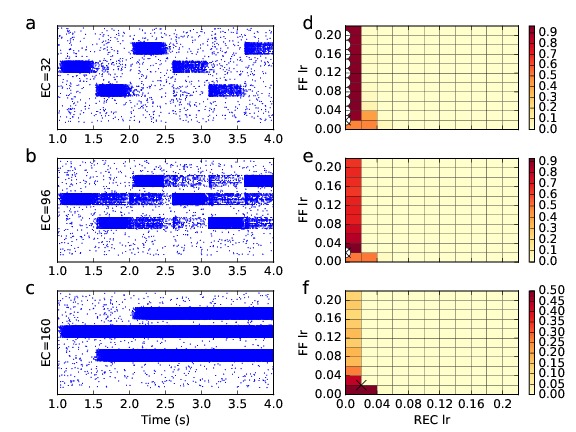}
\par\end{centering}
\caption{Raster plots for encoding (at $1\protect\s$) and retrieval (at $2.5\protect\s$)
of $3$ stimuli, and colormaps for $SNR$ given range of FF and REC
learning rates. (a) Raster plot for simulation with NG, $32$ EC neurons,
and FF and REC learning rates at $0.02$ and $0.0$ respectively.
(b) Raster plot for simulation with NG, $96$ EC neurons, and FF and
REC learning rates as above. (c) Raster plot for simulation with NG,
$160$ EC neurons, and FF and REC learning rates at $0.02$ and $0.02$
respectively. (d) Colormap for $SNR$ with given range of FF and REC
learning rates, for simulations with NG and $32$ EC neurons. White
circle markers denote learning rates with $SNR$ values above $90$
percentile of the maximum $SNR$ value. Black cross marker denotes
learning rates for raster plot to its left. (e) Colormap for $SNR$
with NG and $96$ EC neurons, markers as above. (f) Colormap for $SNR$
with NG and $160$ EC neurons, markers as above.\label{fig:1pl3loop}}
\end{figure}

\begin{figure}
\begin{centering}
\includegraphics[width=0.5\textwidth]{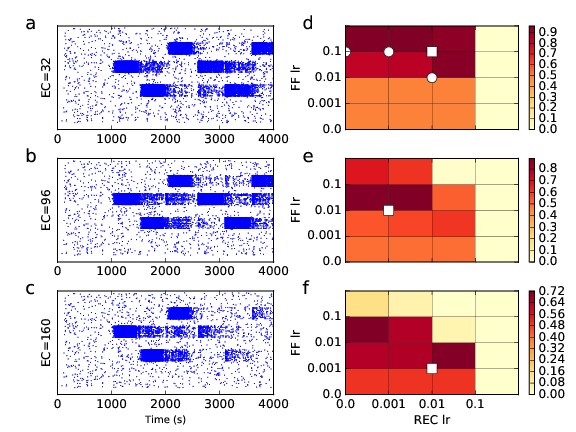}
\par\end{centering}
\caption{Raster plots for encoding (at $1\protect\s$) and retrieval (at $2.5\protect\s$)
of $3$ stimuli, and colormaps for $SNR$ given range of FF and REC
learning rates. (a) Raster plot for simulation with NG, $32$ EC neurons,
and FF and REC learning rates at $0.1$ and $0.01$ respectively.
(b) Raster plot for simulation with NG, $96$ EC neurons, and FF and
REC learning rates at $0.01$ and $0.001$ respectively. (c) Raster
plot for simulation with NG, $160$ EC neurons, and FF and REC learning
rates at $0.001$ and $0.01$ respectively. (d) Colormap for $SNR$
with given range of FF and REC learning rates, for simulations with
NG and $32$ EC neurons. White circle markers denote learning rates
with $SNR$ values above $90$ percentile of the maximum $SNR$ value.
White square marker denotes learning rates for raster plot to its
left. (e) Colormap for $SNR$ with NG and $96$ EC neurons. White
markers as above. (f) Colormap for $SNR$ with NG and $160$ EC neurons.
White markers as above.\label{fig:1pl3loop_smallerrange}}
\end{figure}

Here, by increasing the number of stimuli to $3$, we further investigate
how the full plasticity model would perform given the same set of
parameters as in \prettyref{subsec:Comparison-of-plasticity}. As
observed, in the case of $32$ EC neurons per stimulus, retrieval
of memories encoded achieves good $SNR$ (see \prettyref{fig:1pl3loop}a,d)
for the case of FF plasticity only. When REC plasticity is included,
spiking activities from memory encoding phase persist into the retrieval
phase, adversely affecting the $SNR$ (see \prettyref{fig:1pl3loop}b,c).
This observation can also be made in the $SNR$ colormaps for the
learning rates under all EC population sizes $\{32,96,160\}$. The
space of FF learning rates with better retrieval is larger in the
case of $32$ EC neurons per stimulus as compared to larger ($96$
and $160$) number of EC neurons per stimulus (see \prettyref{fig:1pl3loop}d-f).
This is mainly due to the fact that with more EC neurons per stimulus,
there are more FF synaptic connections on the EC-CA3 pathway potentiated
which increase overall excitation of the CA3 population, leading to
poorer $SNR$ overall. REC plasticity under all EC population sizes
lead to persistent spiking activities of encoding CA3 neurons. Hence,
scanning over the $SNR$ results in the current range of learning
rates (lr), we observe that using a logarithmic scale for the learning
rates ($lr=\{0.0,0.001,0.01,0.1\}$) and with a smaller range, is
sufficient and capture learning rates that give better $SNR$, as
show by the raster plots in \prettyref{fig:1pl3loop_smallerrange}a-c,
and $SNR$ colormaps in \prettyref{fig:1pl3loop_smallerrange}d-f.
We further observe that for high REC $lr=0.1$, spiking activities
in the encoding phase persist into the retrieval phase for all EC
sizes (\prettyref{fig:1pl3loop_smallerrange}d-f), regardless of FF
learning rates. This also happens for high FF $lr=0.1$, when the
EC size per stimulus is $160$ (\prettyref{fig:1pl3loop_smallerrange}c,f),
regardless of REC learning rates. To discover what the optimal set
of parameters are for memory encoding and retrieval, we repeat the
above simulations using $40$ stimuli (fully utilizing the whole DG
network), with both FF and REC $lr=\{0.0,0.001,0.01,0.1\}$.

\subsection*{Determining the best parameters\label{subsec:Determining-the-best}}

\begin{figure}
\begin{centering}
\includegraphics[width=0.5\textwidth]{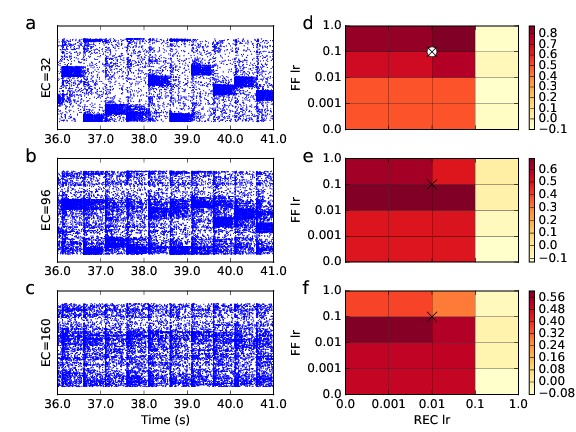}
\par\end{centering}
\caption{Raster plots for encoding and retrieval of $40$ stimuli (retrieval
of last $10$ stimuli shown), and colormaps for $SNR$ with given
range of FF and REC learning rates, with NG. (a) Raster plot for simulation
with $32$ EC neurons, NG and with FF and REC learning rates at $0.1$
and $0.01$ respectively. (b) Raster plot as above, for $96$ EC neurons.
(c) Raster plot as above, for $160$ EC neurons. (d) Colormap for
$SNR$ with given range of FF and REC learning rates, for simulations
with NG and $32$ EC neurons. White markers denote learning rates
with $SNR$ values above $95$ percentile of the maximum $SNR$ value.
Black cross marker denotes learning rates for raster plot to its left.
(e) Colormap as above, for $96$ EC neurons. Marker as above. (f)
Colormap as above, for $160$ EC neurons. Marker as above.\label{fig:40loop}}

\end{figure}

\begin{figure}
\begin{centering}
\includegraphics[width=0.5\textwidth]{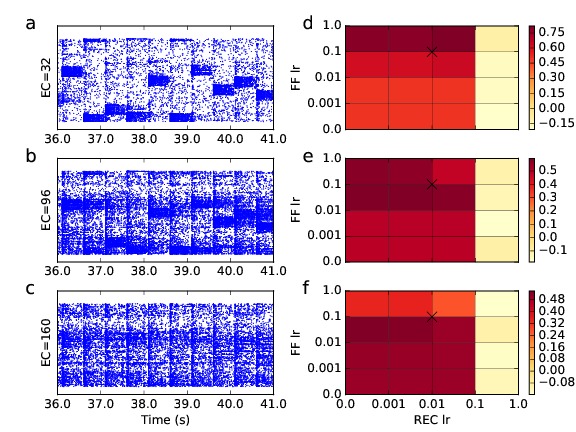}
\par\end{centering}
\caption{Raster plots for encoding and retrieval of $40$ stimuli (retrieval
of last $10$ stimuli shown), and colormaps for $SNR$ with given
range of FF and REC learning rates, without NG. (a) Raster plot for
simulation with $32$ EC neurons, without NG and with FF and REC learning
rates at $0.1$ and $0.01$ respectively. (b) Raster plot as above,
for $96$ EC neurons. (c) Raster plot as above, for $160$ EC neurons.
(d) Colormap for $SNR$ with given range of FF and REC learning rates,
for simulations with NG and $32$ EC neurons. Black cross marker denotes
learning rates for raster plot to its left. (e) Colormap as above,
for $96$ EC neurons. (f) Colormap as above, for $160$ EC neurons.\label{fig:Raster-plots-for40-woNG}}
\end{figure}

In this set of simulations, the learning rates used are $\{0.0,0.001,0.01,0.1\}$.
As shown in \prettyref{fig:40loop}a-c, increased number of EC neurons
per stimulus not only increase background activities, but is also
more prone to triggering spiking activities in other CA3 memory engrams.
This is because the EC neurons are randomly picked and maybe involved
in several memory encoding. Hence, during retrieval, the increased
overlap of EC neurons across memories would also trigger spiking activities
in the other memory engrams in CA3. The $SNR$ also deteriorate with
the number of stimuli (compare $3$ and $40$ stimuli). The $SNR$
colormaps for the different number of EC neurons with NG are shown
in \prettyref{fig:40loop}d-f. The corresponding figure for simulation
settings without neurogenesis is shown in \prettyref{fig:Raster-plots-for40-woNG}.
We further notice that memory retrieval is consistently better for
the case of NG than without and that memory retrieval is better for
less EC neurons (compare \prettyref{fig:40loop} and \prettyref{fig:Raster-plots-for40-woNG},
see also \prettyref{tab:Maximum--for}). For the case of $32$ EC
neurons per stimulus, FF and REC plasticity give better $SNR$ than
just FF alone. But these two modes of plasticity give similar $SNR$
results for more EC neurons per stimulus. While previously, with $32$
EC neurons, both signal and noise neurons are more numerous in the
FF+REC case than the FF case (with proportionally more signal neurons),
resulting in a higher $SNR$, this effect is drowned out by the increased
background noise ($96$ and $160$ EC neurons per stimulus, see \prettyref{tab:Maximum--for})
whereby the maximal $SNR$ for higher number of EC neurons is obtained
at higher spike count. We further note that lower learning rates give
the maximum $SNR$ for more EC neurons. We have also run further simulations
such that 1) FF learning rate $=\{1.0\}$ and REC learning rate $=\{0.0,0.001,0.01,0.1\}$
and 2) FF learning rate $=\{0.0\}$ and REC learning rate $=\{0.001,0.01,0.1,1.0\}$
(not shown). Both set of simulation do not yield better $SNR$ than
those above. In particular, for the second set of simulations, whereby
FF learning rate $=0.0$, persistent spiking activities such as those
in \prettyref{fig:1pl3loop}b occur at large REC learning rates, effectively
drowning out spiking activities of retrieved memories. The learning
rates (both FF and FF+REC) for maximum $SNR$ are later used for simulations
of up to $200$ stimuli (those in red in \prettyref{tab:Maximum--for}),
loading the DG up to $5x$ its designated capacity.

\begin{table*}
\begin{centering}
{\scriptsize{}}%
\begin{tabular}{|c|c|c|c|c|c|c|c|c|c|c|c|c|}
\hline 
 & \multicolumn{6}{c|}{{\scriptsize{}With NG}} & \multicolumn{6}{c|}{{\scriptsize{}Without NG}}\tabularnewline
\hline 
\hline 
 & \multicolumn{3}{c|}{{\scriptsize{}FF}} & \multicolumn{3}{c|}{{\scriptsize{}FF+REC}} & \multicolumn{3}{c|}{{\scriptsize{}FF}} & \multicolumn{3}{c|}{{\scriptsize{}FF+REC}}\tabularnewline
\hline 
{\scriptsize{}$32$} & \textbf{\textcolor{red}{\scriptsize{}$0.83$}} & \textbf{\textcolor{red}{\scriptsize{}$75,10(3)$}} & \textbf{\textcolor{red}{\scriptsize{}$0.1,0.0$}} & \textcolor{red}{\scriptsize{}$0.87${*}} & \textcolor{red}{\scriptsize{}$96,21(3)$} & \textcolor{red}{\scriptsize{}$0.1,0.01$} & {\scriptsize{}$0.80$} & {\scriptsize{}$75,15(3)$} & {\scriptsize{}$0.1,0.0$} & {\scriptsize{}$0.82$} & {\scriptsize{}$85,21(3)$} & {\scriptsize{}$0.1,0.01$}\tabularnewline
\hline 
{\scriptsize{}$96$} & \textbf{\textcolor{red}{\scriptsize{}$0.68$}} & \textbf{\textcolor{red}{\scriptsize{}$54,18(4)$}} & \textbf{\textcolor{red}{\scriptsize{}$0.01,0.0$}} & \textcolor{red}{\scriptsize{}$0.68$} & \textcolor{red}{\scriptsize{}$60,23(4)$} & \textcolor{red}{\scriptsize{}$0.01,0.01$} & {\scriptsize{}$0.59$} & {\scriptsize{}$34,15(4)$} & {\scriptsize{}$0.01,0.0$} & {\scriptsize{}$0.59$} & {\scriptsize{}$33,14(4)$} & {\scriptsize{}$0.01,0.001$}\tabularnewline
\hline 
{\scriptsize{}$160$} & {\scriptsize{}$0.60$} & {\scriptsize{}$43,23(6)$} & {\scriptsize{}$0.01,0.0$} & {\scriptsize{}$0.59$} & {\scriptsize{}$42,24(6)$} & {\scriptsize{}$0.01,0.001$} & {\scriptsize{}$0.54$} & {\scriptsize{}$25,17(7)$} & {\scriptsize{}$0.01,0.0$} & {\scriptsize{}$0.53$} & {\scriptsize{}$21,15(8)$} & {\scriptsize{}$0.01,0.001$}\tabularnewline
\hline 
\end{tabular}
\par\end{centering}{\scriptsize \par}
\caption{Maximum $SNR$ for simulations with $40$ stimuli. The results are
tabulated for the different number of EC neurons, and the case of
with and without NG. A maximum $SNR$ (first column) is chosen for
the case of FF (EC-CA3) plasticity only, and FF+REC (CA3-CA3) plasticity.
In the second column ($x,y(z)$), $x$ denotes the number of CA3 neurons
spiking during retrieval that are in the set of stimulated CA3 neurons
during encoding (signal), while $y$ denotes the number of CA3 neurons
spiking that are not in the set of stimulated CA3 neurons (noise)
and $z$ denotes the spike count used to calculate the optimal $SNR$.
In the third column ($x,y$), $x$ denotes the learning rate for FF
plasticity, while $y$ denotes the learning rate for the REC plasticity.
{*} denotes the maximum $SNR$.\label{tab:Maximum--for} }
\end{table*}

\subsection{Effect of noise on SNR \label{subsec:Effect-of-noise}}

\begin{figure}
\begin{centering}
\includegraphics[width=0.5\textwidth]{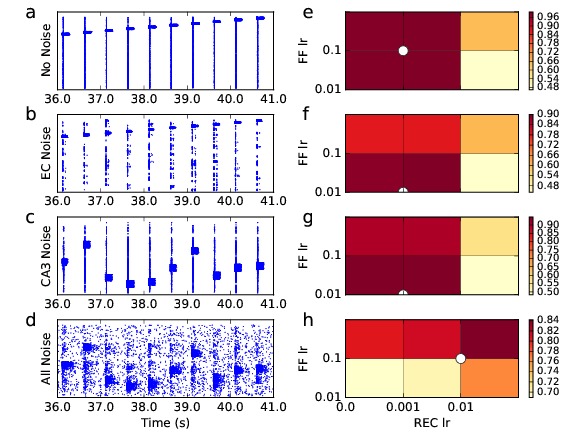}
\par\end{centering}
\caption{Raster plots for encoding and retrieval of $40$ stimuli (retrieval
of last $10$ stimuli shown), and colormaps for $SNR$ with given
range of FF and REC learning rates, with different noise sources.
(a) Raster plot for simulation with $32$ EC neurons, without any
noise and with FF and REC learning rates at $0.1$ and $0.001$ respectively.
(b) Raster plot for $32$ EC neurons, with EC noise only and with
FF and REC learning rates as above. (c) Raster plot for $32$ EC neurons,
with CA3 noise only and with FF and REC learning rates as above. (d)
Raster plot for $32$ EC neurons, with all sources of noise and with
FF and REC learning rates as above. (e) Colormap for $SNR$ with given
range of FF and REC learning rates, for simulations with $32$ EC
neurons and no noise. White marker denotes learning rates for raster
plot to its left, which also gives the maximum $SNR$. (f) Colormap
and marker as above, for EC noise only. (g) Colormap and marker as
above, for CA3 noise only. (h) Colormap and marker as above, for all
sources of noise.\label{fig:Raster-plots-for-noise}}

\end{figure}

In the original network setup, there are $4$ main sources of noise,
namely, 
\begin{enumerate}
\item DG and CA3 (both excitatory and inhibitory) networks receive both
Poisson synaptic inputs and current input so as to be in the fluctuation
driven regime, with a mean firing rate of $0.3\sps$ when idle (Input
noise) 
\item DG neurons have a spiking threshold of $50\mV$ when not encoding
new memories; hence they may still spike, introducing noise into the
network (DG noise)
\item EC neurons are randomly selected to encode each new memory; hence
one EC neuron may be involved in more than one memory. During retrieval,
they may then trigger several memory engrams in the CA3 (EC noise)
\item CA3 neurons are randomly selected to encode each new memory. Hence,
they may also overlap with previously encoded memories. During retrieval
of a memory engram, it may trigger spiking activities in another overlapping
memory engram (CA3 noise)
\end{enumerate}
To investigate the effect of the above noise on the $SNR$, we systematically
switch them off, by
\begin{enumerate}
\item Not introducing Poisson and current inputs into the network
\item Setting DG neurons spiking threshold to $5000\mV$ when not encoding
memories
\item Modifying size of the EC network to $\{1280,3840,6400\}$, such that
all $40$ memories use a new block of $\{32,96,160\}$ EC neurons
respectively
\item Changing the size of each memory engram in the CA3 to $80$ neurons,
and using each such block for encoding a new memory without overlap
\end{enumerate}
There are therefore in total $16$ ($4^{2}$) different noise settings.
We run simulations of $40$ memories, for EC neurons per stimulus$=\{32,96,160\}$,
with NG and $FF\thinspace lr=\{0.001,0.01,0.1\}$ and $REC\thinspace lr=\{0.0,0.001,0.01\}$.
To illustrate the effect of these different noise, we show only the
raster plot and corresponding colormap for the case of $32$ EC neurons
per stimulus,
\begin{itemize}
\item No noise (\prettyref{fig:Raster-plots-for-noise}a,e). The maximum
$SNR$ is $0.99$.
\item Only EC noise (\prettyref{fig:Raster-plots-for-noise}b,f). The maximum
$SNR$ is $0.9$.
\item Only CA3 noise (\prettyref{fig:Raster-plots-for-noise}c,g). The maximum
$SNR$ is $0.95$.
\item All sources of noise present (\prettyref{fig:Raster-plots-for-noise}d,h).
The maximum $SNR$ is $0.84$.
\end{itemize}
DG noise has the least effect on the $SNR$; hence it is not shown.
For all levels of noise, REC plasticity always enhances the $SNR$.

\begin{figure}
\begin{centering}
\includegraphics[width=0.5\textwidth]{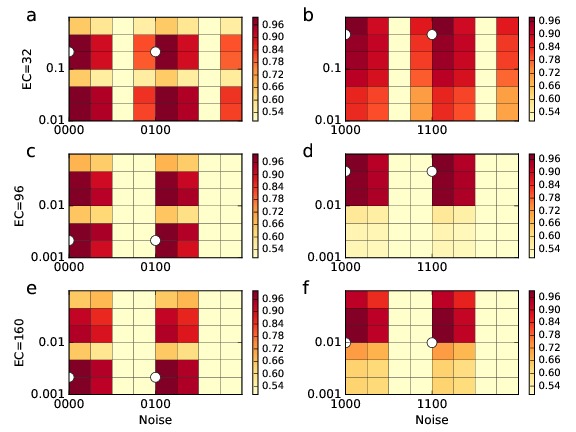}
\par\end{centering}
\caption{Colormaps of $SNR$ for encoding and retrieval of $40$ stimuli with
given range of FF and REC learning rates, and different noise sources.
The X-axis refers to the different noise sources, with the $16$ different
combinations denoted using binary numbers (highest place value denotes
noise source $1$ and so forth). The Y-axis denotes the different
learning rates starting at the origin from $FF\thinspace lr=0.001$
and $REC\thinspace lr=0$ and ending with $FF\thinspace lr=0.01$
and $REC\thinspace lr=0.01$. (a,b) Colormaps for simulation with
$32$ EC neurons, without and with input noise respectively. White
markers denote learning rates and noise combination that give the
maximum $SNR$ for the present parameter range. (c,d) Colormaps for
simulation with $96$ EC neurons, without and with input noise respectively.
White markers as above. (e,f) Colormaps for simulation with $160$
EC neurons, without and with input noise respectively. White markers
as above. The markers in (e) denote the maximum $SNR$ ($1.0)$ for
all settings.\label{fig:noise_colormap}}

\end{figure}

A summary of the $SNR$ results are shown in \prettyref{fig:noise_colormap}.
We observe that good $SNR$ is obtained when all $4$ sources of noise
are present only for the case of $32$ EC neurons (\prettyref{fig:noise_colormap}b,d,f).
In the case whereby EC neurons per stimulus $=\{96,160\}$, we note
that perfect retrieval ($SNR=1.0$) is obtained for the case of $160$
EC neurons per stimulus, $FF\thinspace lr=0.001$ and $REC\thinspace lr=0.001$,
and without noise source $1,3$ and $4$ (\prettyref{fig:noise_colormap}e).
DG threshold ($50\mV$ or $5000\mV$) when not encoding memories does
not matter. We further note that when the network is in the fluctuating
regime (with input noise) and with more EC neurons per stimulus ($\{96,160\}$),
good $SNR$ is obtained with a higher $FF\thinspace lr$ ($0.01$),
as shown by \ref{fig:noise_colormap}d,f. Other than for \ref{fig:noise_colormap}f,
the best $SNR$ for all other simulation settings is obtained when
$REC\thinspace lr$ is non-zero. Hence, in contrast to results summarized
in \ref{tab:Maximum--for}, for all EC sizes, when there is no noise
in the network, FF+REC plasticity give better retrieval results than
just FF plasticity alone, i.e. pattern completion. This clearly illustrates
that in the noisy network (all noises present), the noise is amplified
by the REC plasticity, and with larger EC sizes ($96$ and $160$),
the increased network noise cancels out the pattern completion effect
of REC plasticity, resulting in $SNR$ on par with FF plasticity alone.

\subsection*{Robustness of results in DG overloading\label{subsec:Robustness-of-results}}

\begin{figure}
\begin{centering}
\includegraphics[width=0.5\textwidth]{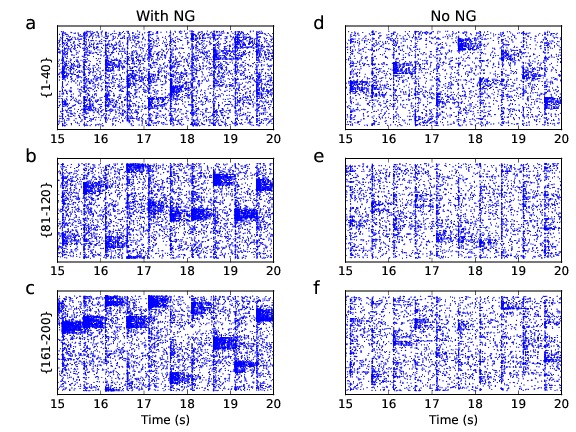}
\par\end{centering}
\caption{Raster plots for retrieval showing last $10$ retrieved memories.
(a-c) Raster plots for retrievals ($\{1-40\}$,$\{81-120\}$ and $\{161-200\}$)
with NG, $32$ EC neurons, and with FF and REC learning rates at $0.1$
and $0.01$ respectively. (d-f) Raster plots as above, for simulations
without NG.\label{fig:Raster-plots-for-200}}

\end{figure}

For this set of simulations, we select simulation settings in \prettyref{tab:Maximum--for}
with red color font for overloading (with NG). Firstly, we tried simulations
each with $80$, $120$, $160$ or $200$ numbers of stimuli and then
retrieved the last $40$ encoded memories for each simulation. Next,
we tried encoding $200$ memories and then retrieved them in batches
of $40$ memories each: $\{1-40\},\{41-80\},\{81-120\},\{121-160\},\{161-200\}$.
Third, the second set of simulations are repeated with partial retrieval
such that only $0.25$, $0.5$ or $0.75$ of encoding EC neurons are
stimulated during retrieval. In general, the observations made agree
with those summarized in \prettyref{tab:Maximum--for}. The $SNR$
results described below are for the case whereby there are $32$ EC
neurons per stimulus, $FF\thinspace lr=0.1$ and $REC\thinspace lr=0.01$,
which gives the best $SNR$. In the first simulations, more memories
encoded result in poorer retrieval. For example, the $SNR$ for retrieving
the last $40$ memories for $80$, $120$, $160$ and $200$ numbers
of stimuli are respectively $0.86$, $0.84$, $0.81$ and $0.76$.
The $SNR$ for the retrieved memory batches in the second simulations
($\{1-40\},\{41-80\},\{81-120\},\{121-160\},\{161-200\}$) are respectively
$0.51$, $0.57$, $0.63$, $0.71$ and $0.76$ (\prettyref{fig:Raster-plots-for-200}a-c).
Hence, older memories are ``forgotten''. The $SNR$ for the third
set of simulations (partial retrieval of encoded memories $\{161-200\}$)
are respectively $0.72$, $0.66$ and $0.55$ for $0.75$, $0.5$
and $0.25$ of encoding EC neurons stimulated during retrieval. Hence,
with only $0.75$ of EC neurons activated during retrieval, the retrieval
results are comparable ($SNR$ $0.76$ and $0.72$). The $SNR$ for
the partial retrieval of encoded memories $\{41-80\}$ are respectively
$0.55$, $0.51$ and $0.5$ for $0.75$, $0.5$ and $0.25$ of encoding
EC neurons stimulated during retrieval. Hence, partial retrieval can
be achieved but only for the more recent memories. In the above simulations,
$32$ stimulating EC neurons give better $SNR$ results compared to
$96$ EC neurons, and FF+REC plasticity perform better than just FF
plasticity alone. 

The above results are for the case of NG (reduced spiking threshold
for encoding DG neurons during stimulation and randomly selected DG
neurons are recycled, i.e. previous synaptic connections are set to
$0\ns$, after the DG network is fully utilized). In the case of without
neurogenesis (spiking threshold for encoding DG neurons remained the
same and selected DG neurons keep their previous synaptic connections),
the second set of simulations are repeated. The $SNR$ results for
the case where there are $32$ EC neurons per stimulus, $FF\thinspace lr=0.1$
and $REC\thinspace lr=0.01$, and memory batches $\{1-40\},\{41-80\},\{81-120\},\{121-160\},\{161-200\}$
are respectively $0.54,0.54,0.53,0.52,0.51$ (\prettyref{fig:Raster-plots-for-200}d-f).
Without neurogenesis, as expected, DG neurons are involved in encoding
of many more memories, and low $SNR$ is evenly spread across all
stimuli, but with slightly better $SNR$ in retrieval of earlier batches,
a trend different from the case of NG. We have also repeated the retrieval
of the last $40$ encoded memories ($\{161-200\}$) in the second
set of simulation for both with and without NG with a $20000\ms$
interval between encoding and retrieval. The $SNR$ are respectively
also $0.76$ and $0.51$, same as those without the interval. 

\begin{figure}
\begin{centering}
\includegraphics[width=0.5\textwidth]{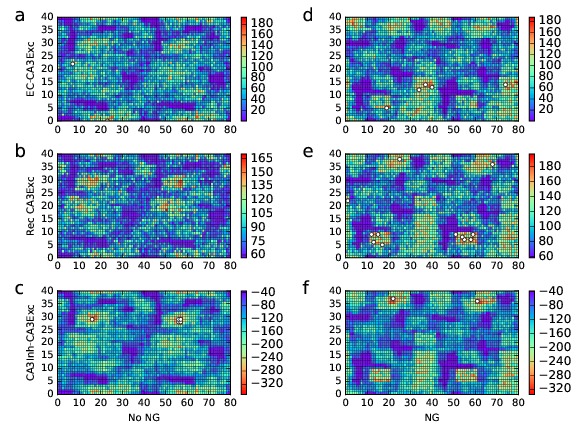}
\par\end{centering}
\caption{Colormaps for sum of weights of synaptic connections from different
network populations (EC, CA3 excitatory and CA3 inhibitory) in the
CA3 excitatory population, after $80$ memories are encoded. (a-c)
Colormaps in the case of no NG for synaptic connections from the EC,
CA3 excitatory (recurrent connections) and CA3 inhibitory populations
respectively. White markers denote neuron with sum of weights above
$90$ percentile of the maximum sum (as of both with and without NG).
(d-f) Colormaps and markers as above, for the case with NG.\label{fig:Colormaps-for-80}}

\end{figure}

\begin{figure}
\begin{centering}
\includegraphics[width=0.5\textwidth]{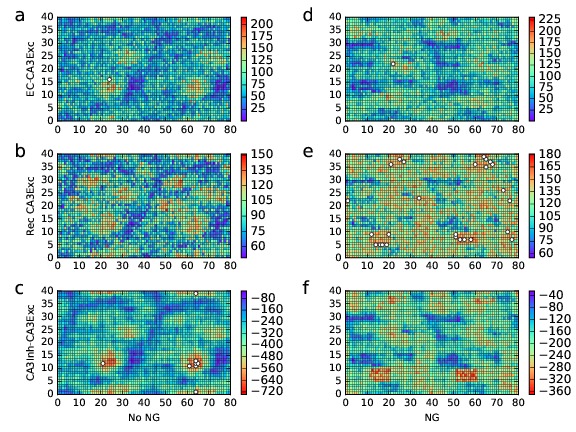}
\par\end{centering}
\caption{Colormaps for sum of weights of synaptic connections from different
network populations (EC, CA3 excitatory and CA3 inhibitory) in the
CA3 excitatory population, after $200$ memories are encoded. (a-c)
Colormaps in the case of no NG for synaptic connections from the EC,
CA3 excitatory (recurrent connections) and CA3 inhibitory populations
respectively. White markers denote neuron with sum of weights above
$90$ percentile of the maximum sum (as of both with and without NG).
(d-f) Colormaps and markers as above, for the case with NG.\label{fig:Colormaps-for-200}}

\end{figure}

\begin{figure}
\begin{centering}
\includegraphics[width=0.5\textwidth]{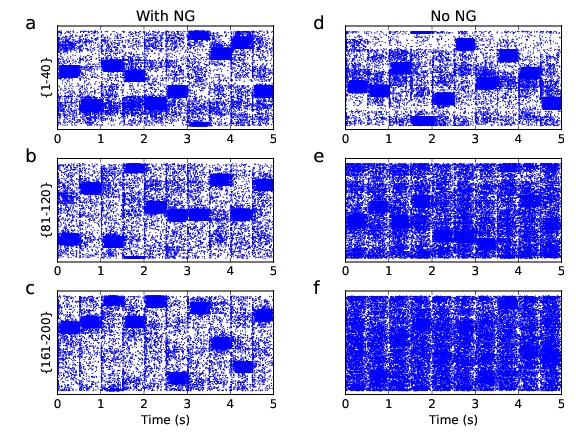}
\par\end{centering}
\caption{Raster plots for encoding showing last $10$ encoded memories. (a-c)
Raster plots for encoding ($\{1-40\}$,$\{81-120\}$ and $\{161-200\}$)
with NG, $32$ EC neurons, and with FF and REC learning rates at $0.1$
and $0.01$ respectively. (d-f) Raster plots as above, for simulations
without NG.\label{fig:Raster-plots-for-encoding}}

\end{figure}

To investigate why memory retrieval for large number of memories are
so different for the case of with and without neurogenesis, we look
at the sums of weights of the different incoming synaptic connections
in the CA3 excitatory population after encoding of $80$ \prettyref{fig:Colormaps-for-80}
and $200$ \prettyref{fig:Colormaps-for-200} memories respectively.
Firstly, we observe that there are more clusters of large sums of
CA3 excitatory synapses in the case of NG than without for both after
$80$ (\prettyref{fig:Colormaps-for-80}b,e) and $200$ (\prettyref{fig:Colormaps-for-200}b,e)
encoded memories. Next, we observe that sums of inhibitory connections
have increased much more in the case of no NG (compare \prettyref{fig:Colormaps-for-80}c
and \prettyref{fig:Colormaps-for-200}c) than when there is NG (compare
\prettyref{fig:Colormaps-for-80}f and \prettyref{fig:Colormaps-for-200}f)
from $80$ to $200$ encoded memories. This is due to the fact that
when new memories are encoded in the no NG case, EC-DG and DG-CA3
excitatory synaptic connections are not set to $0\ns$ when DG neurons
are reused. Hence for each new memory to be encoded, in the no NG
case, increasingly more CA3 excitatory neurons are activated, in contrast,
about the same number of CA3 neurons are activated in the NG case
during encoding of all $200$ memories (\prettyref{fig:Raster-plots-for-encoding}).
This leads to a much larger increase of noisy inputs in the CA3 excitatory
population (compared to the case with NG), which must be counter-balanced
by a larger increase in CA3 inhibitory connections so as to keep the
CA3 populations stable. This large increase in network inhibition
plays an active role in abolishing earlier encoded memories. This
more active recruitment of inhibitory plasticity in the CA3 networks
(in the no NG case) also leads to decrease in recurrent excitation
(\prettyref{fig:Colormaps-for-200}b). We further note that overall,
for the CA3 population with NG, there is a bigger increase in the
excitatory connections (EC-CA3 excitatory and CA3 excitatory-CA3 excitatory),
than without NG, as more memories are encoded (from $80$ to $200$
memories, compare \prettyref{fig:Colormaps-for-80}a,b,d,e and \prettyref{fig:Colormaps-for-200}a,b,d,e).
The above observations made using the colormaps agree with the synaptic
weight histograms in \prettyref{sec:Supplementary-materials}.

Hence, the above observations illustrate that, in the case of no NG,
while the different plasticity mechanisms are able to maintain the
stability of the network (a global network property), with more CA3
excitatory neurons activated with each new memory to be encoded, inhibitory
plasticity loses its specificity. This results in the abolition of
memories encoded in earlier batches. The ``forgetting'' observed
in the NG and no NG cases are however due to different mechanisms.
In the NG case, it is a direct consequence of deletion of connections.
In the no NG case, it is due to increased inhibition. These could
perhaps help to explain the different trends in the $SNR$ for the
different batches.

\section{Discussion\label{sec:Discussion}}

Using the network and simulation settings as described in \ref{sect_methods},
we first investigate how the different models of plasticity perform
under a wide range of parameters. Next, using the selected plasticity
model (full plasticity model), we increase the number of memories
encoded to $3$, and as a result, arrive at a narrower range (logarithmic
scale) of learning rates. Third, we further increase the number of
memories to $40$, so as to fully utilize the DG, and from the simulations
performed, show that for all learning rates ($FF\thinspace lr=\{0.0,0.001,0.01,0.1\}$
and $REC\thinspace lr=\{0.0,0.001,0.01,0.1\}$) and number of EC neurons
per stimulus ($\{32,96,160\}$), memory retrieval is better when there
is NG. Also, for the case when there are $32$ EC neurons per stimulus,
REC plasticity enhances the $SNR$, as compare to just FF plasticity,
illustrating the importance of REC plasticity for pattern completion.
However, when the network has more noise (for instance, when number
of EC neurons per stimulus $=\{96,160\}$), REC plasticity leads to
co-activation of overlapping memory engrams in the CA3, adversely
affecting the $SNR$, resulting in poorer retrieval results compared
to just FF plasticity alone. This is further shown in the fourth set
of simulations whereby FF+REC plasticity result in better retrieval
for the case of $160$ EC neurons per stimulus when there is no noise
in the network, as compared to just FF plasticity alone. There is
therefore a fine balance between when recurrent plasticity plays an
enhancing or diminishing role in memory retrieval. Finally, we perform
simulations whereby up to $200$ memories are encoded and then retrieved.
We show that with $200$ memories, NG enables the retrieval of the
most recent set of encoded memories, while without NG, the signal
strength of retrieved memories is roughly that of background noise
($SNR\approx0.5$). This is mainly due to the fact that without NG,
each new stimulus will excite ever more excitatory CA3 neurons, which
in turn will result in increased inhibitory plasticity so as to maintain
network stability, which effectively erases the memory encoded (even
most recent ones). While with NG, each stimulus will excite about
the same number of excitatory CA3 neurons as previous stimuli, and
in turn, result in a more targeted (and less) increment in inhibitory
plasticity, which helps to preserve the more recent stimuli encoded.
Through NG, however, older memories encoded are erased through deletion
of EC-DG and DG-CA3 connections. 

Hence in our model, neurogenesis provides the structural plasticity
(creation and deletion of connections) required for encoding and retrieval
of memories when the network is overloaded. Synaptic plasticity, on
its own, does not seem to be sufficient. Even if the EC-DG and DG-CA3
pathways have been plastic, the old connections would only potentiate
with each new encoding memory, and more CA3 neurons would still spike
with each new memory, as shown in the above simulations without neurogenesis.
Specificity of memory engrams is thus lost in the process. In addition,
to maintain network stability, inhibitory plasticity, operating on
the same time-scale (necessarily so), effectively suppresses network
activities, and any memory engram that may have formed is erased.

From our simulations, memory retrieval for up to $200$ encoded stimuli
is achieved even when the plasticity is kept on during the retrieval
phase, despite using a simplified version of the plasticity models
in~\citep{Zenke2015}. Retrieval is also shown to work equally well
when there is an interval of $20000\ms$ between encoding and retrieval.
For the case of both FF+REC plasticity, memory engrams formed using
both the EC-CA3 pathways and REC CA3 connections. This gives better
$SNR$ than just FF plasticity alone, while there is less network
noise. Hence plastic recurrent connection plays a role in pattern
completion (improving memory retrieval especially partial ones), but
also amplifies the noise in the network (retrieving more than what
is encoded); this warrants further studies. While REC plasticity alone
may also encode memories, this occurs at high REC learning rates,
whereby persistent spiking activities dominate the network. Hence,
retrieved memory engrams are effectively drowned out by the persistent
activities, leading to poor $SNR$. The highly REC CA3 network hence
operate best as an associative memory by sharing the task of encoding
memory engrams between both the FF and REC synaptic connections, at
a low noise level.

Memory retrieval deteriorates with increased encoded memories. This
is primarily due to noise in the network, which can be attributed
to $4$ main sources as shown in \prettyref{subsec:Effect-of-noise}.
For EC and CA3 noise, it is observed from \prettyref{tab:Maximum--for}
how noise neurons are generally more in the FF+REC case as compared
to just the FF case. We have systematically remove the above sources
of noise in \prettyref{subsec:Effect-of-noise}, which results in
almost perfect retrieval with $40$ stimuli (no overloading). Such
a setting is however biologically unrealistic as it would imply that
there must be at least as many EC, DG and CA3 neurons as there are
memories. 

In reality, forgetting of older memories is natural. Also, spiking
of other CA3 neurons other than those encoding the particular memory
during retrieval may help to associate several related memories which
maybe useful for downstream processing. Memories in the hippocampus
are also further consolidated in the cortex, of which how they function
as a complete system is still an active field of research. The model
we have studied set up a framework for further extension in computational
studies of the role of the hippocampus in memory encoding and retrieval.

\section{Supplementary materials\label{sec:Supplementary-materials}}

\begin{table}[H]
\begin{raggedright}
\begin{tabular}{|l|c|}
\hline 
\multicolumn{2}{|l|}{Short term plasticity}\tabularnewline
\hline 
\hline 
probability of release & $0.2$\tabularnewline
\hline 
$\tau_{facilitating}$ & $600\ms$\tabularnewline
\hline 
$\tau_{depressing}$ & $400\ms$\tabularnewline
\hline 
\multicolumn{2}{|l|}{Pair STDP}\tabularnewline
\hline 
$\tau_{K^{i}}$ & $20\ms$\tabularnewline
\hline 
$\tau_{K^{j}}$ & $20\ms$\tabularnewline
\hline 
$W_{max}$ & $5.0\ns$\tabularnewline
\hline 
\multicolumn{2}{|l|}{Triplet STDP}\tabularnewline
\hline 
$\tau_{K^{i1}}$ & $20\ms$\tabularnewline
\hline 
$\tau_{K^{i2}}$ & $101\ms$\tabularnewline
\hline 
$\tau_{K^{j1}}$ & $20\ms$\tabularnewline
\hline 
$\tau_{K^{j2}}$ & $100\ms$\tabularnewline
\hline 
$W_{max}$ & $5.0\ns$\tabularnewline
\hline 
\multicolumn{2}{|l|}{Full STDP}\tabularnewline
\hline 
$\tau_{K^{i1}}$ & $20\ms$\tabularnewline
\hline 
$\tau_{K^{i2}}$ & $101\ms$\tabularnewline
\hline 
$\tau_{K^{j1}}$ & $20\ms$\tabularnewline
\hline 
$\tau_{K^{j2}}$ & $100\ms$\tabularnewline
\hline 
$W_{max}$ & $5.0\ns$\tabularnewline
\hline 
$\beta$ & $0.01\ms^{-1}$\tabularnewline
\hline 
$\tilde{W}$ & $0.2\ns$\tabularnewline
\hline 
$\delta$ & $2\exp(-5)$\tabularnewline
\hline 
\multicolumn{2}{|l|}{Inhibitory plasticity}\tabularnewline
\hline 
$\tau_{K^{i}}$ & $20\ms$\tabularnewline
\hline 
$\tau_{K^{j}}$ & $20\ms$\tabularnewline
\hline 
$W_{max}$ & $48.0\ns$\tabularnewline
\hline 
$\alpha$ & $0.02$\tabularnewline
\hline 
\end{tabular}
\par\end{raggedright}
\caption{Plasticity parameters used in all simulations}
\end{table}

\begin{table}[H]
\begin{tabular}{|l|l|}
\hline 
\multicolumn{2}{|l|}{{\scriptsize{}EC (standalone neurons)}}\tabularnewline
\hline 
\hline 
{\scriptsize{}neurons} & {\scriptsize{}$1024$ AIF neurons}\tabularnewline
\hline 
\multicolumn{2}{|l|}{{\scriptsize{}DG (standalone neurons)}}\tabularnewline
\hline 
{\scriptsize{}neurons} & {\scriptsize{}$3200$ AIF neurons}\tabularnewline
\hline 
{\scriptsize{}Poisson excitatory input} & {\scriptsize{}$180\sps$}\tabularnewline
\hline 
{\scriptsize{}Poisson inhibitory input} & {\scriptsize{}$45\sps$}\tabularnewline
\hline 
{\scriptsize{}Current input} & {\scriptsize{}$4.5\pa$}\tabularnewline
\hline 
{\scriptsize{}synaptic weights} & {\scriptsize{}excitatory : $0.1\ns$, inhibitory : $0.44\ns$}\tabularnewline
\hline 
\multicolumn{2}{|l|}{{\scriptsize{}CA3 excitatory (40{*}80 on topology layer)}}\tabularnewline
\hline 
{\scriptsize{}neurons} & {\scriptsize{}$3200$ AIF neurons}\tabularnewline
\hline 
{\scriptsize{}Poisson excitatory input} & {\scriptsize{}$240\sps$}\tabularnewline
\hline 
{\scriptsize{}Poisson inhibitory input} & {\scriptsize{}$60\sps$}\tabularnewline
\hline 
{\scriptsize{}Current input} & {\scriptsize{}$4.5\pa$}\tabularnewline
\hline 
{\scriptsize{}synaptic weights} & {\scriptsize{}excitatory : $0.1\ns$, inhibitory : $0.44\ns$}\tabularnewline
\hline 
\multicolumn{2}{|l|}{{\scriptsize{}CA3 inhibitory (20{*}40) on topology layer}}\tabularnewline
\hline 
{\scriptsize{}neurons} & {\scriptsize{}$800$ AIF neurons}\tabularnewline
\hline 
{\scriptsize{}Poisson excitatory input} & {\scriptsize{}$240\sps$}\tabularnewline
\hline 
{\scriptsize{}Poisson inhibitory input} & {\scriptsize{}$60\sps$}\tabularnewline
\hline 
{\scriptsize{}Current input} & {\scriptsize{}$4.5\pa$}\tabularnewline
\hline 
{\scriptsize{}synaptic weights} & {\scriptsize{}excitatory : $0.1\ns$, inhibitory : $0.44\ns$}\tabularnewline
\hline 
\multicolumn{2}{|l|}{{\scriptsize{}CA3 network recurrent connections}}\tabularnewline
\hline 
{\scriptsize{}CA3 exc-\textgreater{}CA3 exc} & {\scriptsize{}full plasticity; convergent (probability:$0.1875)$}\tabularnewline
\hline 
{\scriptsize{}CA3 exc-\textgreater{}CA3 inh} & {\scriptsize{}short term plasticity; convergent (probability:$0.1875)$ }\tabularnewline
\hline 
{\scriptsize{}CA3 inh-\textgreater{}CA3 exc} & {\scriptsize{}inhibitory plasticity; convergent (probability:$0.1875)$}\tabularnewline
\hline 
{\scriptsize{}CA3 inh-\textgreater{}CA3 inh} & {\scriptsize{}static synapse; convergent (probability:$0.1875)$ }\tabularnewline
\hline 
\multicolumn{2}{|l|}{{\scriptsize{}EC-\textgreater{}CA3 exc connections}}\tabularnewline
\hline 
{\scriptsize{}EC-\textgreater{}CA3 exc connections} & {\scriptsize{}full plasticity; divergent (fixed outdegree:$480)$,
weight:$0\ns$, delay:$0.2\ms$}\tabularnewline
\hline 
\multicolumn{2}{|l|}{{\scriptsize{}Encoding settings}}\tabularnewline
\hline 
{\scriptsize{}Stimulus} & {\scriptsize{}$50$ spikes with $5\ms$ interval onto EC neurons,
weight:$10\ns$}\tabularnewline
\hline 
{\scriptsize{}EC-\textgreater{}DG connections} & {\scriptsize{}Convergent (fixed indegree:$2$), weight:$0.4\ns$}\tabularnewline
\hline 
{\scriptsize{}DC-\textgreater{}CA3 exc connections} & {\scriptsize{}Convergent (fixed indegree:$3$), weight:$0.4\ns$}\tabularnewline
\hline 
\multicolumn{2}{|l|}{{\scriptsize{}Retrieval settings}}\tabularnewline
\hline 
{\scriptsize{}Stimulus} & {\scriptsize{}$50$ spikes with $5\ms$ interval onto EC neurons,
weight:$10\ns$}\tabularnewline
\hline 
{\scriptsize{}EC-\textgreater{}DG connections} & {\scriptsize{}Convergent (fixed indegree:$2$), weight:$0\ns$}\tabularnewline
\hline 
{\scriptsize{}DC-\textgreater{}CA3 exc connections} & {\scriptsize{}Convergent (fixed indegree:$3$), weight:$0\ns$}\tabularnewline
\hline 
\end{tabular}\caption{Generic network setup for encoding and retrieval of memories. Synaptic
delay set at $0.1\protect\ms$ unless otherwise stated. All synapses
(including from Poisson generators to neurons) are static unless otherwise
stated. }

\end{table}

\begin{figure}[H]
\begin{centering}
\includegraphics[width=0.5\textwidth]{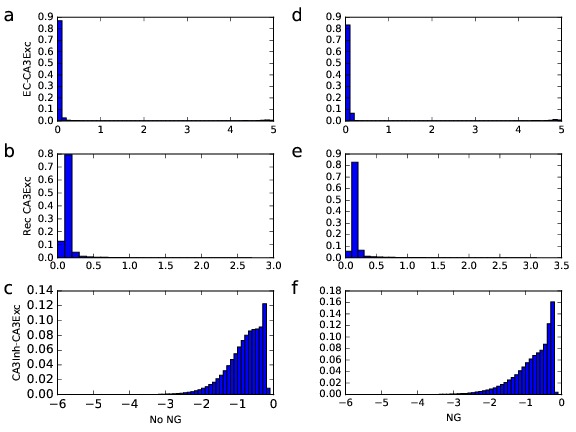}
\par\end{centering}
\caption{Histogram for weights of synaptic connections (EC-CA3 excitatory,
CA3 excitatory-CA3 excitatory, CA3 inhibitory-CA3 excitatory), after
$80$ memories are encoded. (a-c) Histograms in the case of no NG
for the different synaptic connections. (d-f) Histograms as above,
for the case with NG.}
\end{figure}

\begin{figure}[H]
\begin{centering}
\includegraphics[width=0.5\textwidth]{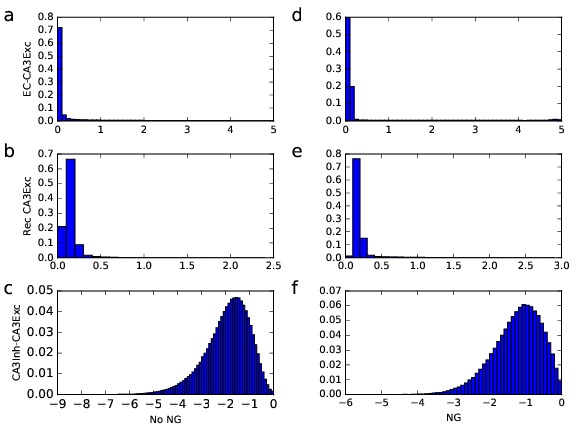}
\par\end{centering}
\caption{Histogram for weights of synaptic connections (EC-CA3 excitatory,
CA3 excitatory-CA3 excitatory, CA3 inhibitory-CA3 excitatory), after
$200$ memories are encoded. (a-c) Histograms in the case of no NG
for the different synaptic connections. (d-f) Histograms as above,
for the case with NG.}

\end{figure}



{\small{}\bibliographystyle{aaai}
\bibliography{nips_2016_bib}
}{\small \par}
\end{document}